\documentclass[eqsecnum,superscriptaddress,showpacs]{revtex4}
\usepackage{graphics}
\usepackage{latexsym}
\usepackage{bm}
\newcommand{\Slash}[1]{{\ooalign{\hfil/\hfil\crcr$#1$}}}
\begin{document}
\title{A Proposal to Measure Photon-Photon Scattering}

\author{Takehisa Fujita}\email{fffujita@phys.cst.nihon-u.ac.jp}
\author{Naohiro Kanda}\email{nkanda@phys.cst.nihon-u.ac.jp}
\affiliation{Department of Physics, Faculty of Science and Technology, 
Nihon University, Tokyo, Japan}

\date{\today}%

\begin{abstract}

We discuss a possibility to measure the photon-photon scattering cross section 
at low energy in a theoretical standpoint. The cross section of photon-photon 
scattering at low energy can be written as 
$ \displaystyle{{d\sigma\over d\Omega} \simeq {\alpha^4\over 
(12\pi)^2 \omega^2} (3+2\cos^2\theta +\cos^4\theta )     } $ with $\omega$ the energy 
of photon. The magnitude of the cross section at $\omega \simeq 1$ eV should be 
$10^{37}$ times larger than the prediction of Heisenberg and Euler who calculated 
the photon scattering by the classical picture of field theory. Due to a difficulty of 
the initial condition of photon-photon reaction process, we propose to first measure  
$\gamma +\gamma \rightarrow e^++e^- $ reaction at a few MeV before measuring 
$\gamma +\gamma \rightarrow \gamma +\gamma $ elastic scattering.

\end{abstract}


\maketitle

\section{Introduction}
It is well known that photon interacts with photon via the box diagrams where 
fermions and anti-fermions are created from the vacuum state. This calculation 
of the Feynman diagrams can be carried out in a straightforward fashion, and one 
can find the cross section of photon-photon scattering in the low energy region 
\cite{kanda} 
$$ {d\sigma\over d\Omega} \simeq {\alpha^4\over 
(12\pi)^2 \omega^2} (3+2\cos^2\theta +\cos^4\theta ) \qquad \ \ \  
({\rm with} \ \ \omega << m)  \eqno{(1.1)}    $$
where $\omega$ and $m$ denote the photon energy and the mass of electron, respectively. 
This is calculated in the center of system of two photons. This cross section is quite 
reasonable in that it is described in terms of the energy of photon only. There appears 
no internal energy scale such as electron mass which should appear in the box diagrams, 
and this is because the cross section  is evaluated in the low energy limit. 

On the other hand, the photon-photon cross section was obtained by the classical 
picture of scattering procedure long time ago, and it is written as \cite{landau}
$$ {d\sigma\over d\Omega}\simeq {139\alpha^4\over (180\pi)^2 m^2} 
\left({\omega\over m} \right)^6 (3+\cos^2\theta )^2   \qquad \ \ \  
({\rm with} \ \ \omega << m)  \eqno{(1.2)}  $$
which was calculated by Heisenberg and Euler in 1936 \cite{euler}, and later 
the result is confirmed by Karplus and Neuman \cite{karp}. Since then, it is 
believed that this photon-photon cross section is the correct one, even though the quantum 
evaluation of the Feynman diagram gives the cross section of eq.(1.1) unless one 
should put some additional but unphysical conditions. In addition, the effective 
Lagrangian method proposed by Heisenberg and Euler for the vacuum polarization effects 
is physically incorrect since it disagrees with the observation that photon is always 
massless \cite{fk}. 

Up to now, the measurements of the photon-photon scattering cross section have been made 
by Moulin et al. \cite{exp1,exp2}, but they found no evidence of the photon-photon 
scattering. However the measurements with the sufficient accuracy must be very difficult  
since photon cannot be at rest but always at the speed of light. Most of the scattering 
experiments are the collision experiment of the incident particle with some fixed targets 
which are basically taken to be at rest. In this respect, the photon-photon scattering 
must be quite new to the conventional experiments, and therefore this experiment must be 
one of the most important experiments in particle physics, 
which is left almost untouched until now. In this sense, the main difficulty must be 
connected to the initial condition of the scattering experiments in which one should 
control the time of photon-photon collision and the focusing of the photon-photon beams. 
In addition, the photon-photon scattering is the reaction process arising from 
the particle nature of photon, in contrast to the wave nature of photon such as 
diffraction or interference phenomena. 

In this respect, we believe that the $\gamma + \gamma \rightarrow e^++e^- $ experiment 
should be first used as the monitor of the initial state condition check of 
the photon-photon scattering experiment. The reaction cross section of 
$\gamma + \gamma \rightarrow e^++e^- $ is 
the same order of magnitude as the Compton scattering when the incident photon energy is 
larger than a few MeV. Therefore, it is crucial that the experimental setup should be 
able to reproduce the cross section of $\gamma + \gamma \rightarrow e^++e^- $ process. 
Even though the photon-photon cross section should be smaller than the $\gamma + \gamma \rightarrow e^++e^- $ by several order of magnitudes in a few MeV region, it should be 
possible to measure the photon-photon elastic cross section once the problem of 
the initial condition is resolved. We believe that the measurement of two photons should 
not be very difficult indeed, even though a very small number.

\section{Qualitative Behavior of Loop Diagrams}
Here, we should clarify why the photon-photon cross section should have the shape 
of eq.(1.1) which does not depend on the ratio of $\left({\omega\over m}\right)^2$ 
in contrast to eq.(1.2). In order to understand the situation clearly, we first compare 
the box diagram of the photon-photon scattering with the triangle diagram 
of $\pi^0 \rightarrow 2  \gamma $ since there is a good similarity between them. 
Namely, both diagrams contain one loop of fermions. 

\subsection{$\pi^0 \rightarrow 2  \gamma $ Decay}
Now, the T-matrix of $\pi^0 \rightarrow 2  \gamma $ can be evaluated to be
$$ T_{\pi^0 \rightarrow 2 \gamma} \simeq e^2 g \int {d^4p\over (2\pi)^4} {\rm Tr} 
\left[(\gamma \epsilon_1) {1\over p \llap/-M +i\varepsilon } 
(\gamma \epsilon_2) {1\over p \llap/-{k \llap/}_2-M +i\varepsilon }\gamma_5 
{1\over p \llap/+{k \llap/}_1-M+i\varepsilon  }  \right] 
\simeq { e^2g\over M} \varepsilon_{\mu \nu \rho \sigma} \epsilon_1^\mu \epsilon_2^\nu
k_1^\rho k_2^\sigma   \eqno{(2.1)} $$
where $M$ denotes the nucleon mass, and $g$ is the coupling constant of $\pi N$ 
interaction as described by ${\cal L}_I=ig\bar{\psi}\gamma_5 \psi \varphi $. 
The energy of two photons can be written as 
$ \omega_1 =|\bm{k}_1|, \ \ \omega_2 =|\bm{k}_2| $. 
Under the condition of $ M>> \omega_1, \omega_2 $, one finds the leading behavior of 
the finite terms, apart from some numerical factors
$$ T_{\pi^0 \rightarrow 2 \gamma} \sim  e^2g  \left({\mu^2\over M}\right) 
\left(1 +O \left({\mu\over M}\right)^2+.. \right)  \eqno{(2.2)}  $$
where $\mu$ denotes the mass of $\pi^0 $. It should be noted that the apparent 
divergences of the T-matrix are kinematically cancelled out in an exact fashion, and 
thus it is not due to the regularization. Even though this problem is well explained 
in detail in the textbook of Nishijima \cite{nishijima}, it may be better to make 
a comment on the relation between the  $\pi^0 \rightarrow 2  \gamma $ and the chiral 
anomaly, in order to avoid any confusions which are sometimes found in the literatures. 
If one calculates the  $\pi^0 \rightarrow 2  \gamma $ case, then one sees immediately 
that there is no divergence. However, educated people may invoke a triangle anomaly 
in which there is an apparent linear divergence when the vertex is the axial vector 
current, instead of pseudoscalar interaction. However, this process, the origin of 
the chiral anomaly, has no divergence either, if one calculates the Feynman diagrams 
properly, without referring to the textbook description. Therefore, one can convince 
oneself that any physically observable processes have no divergence at all, and this is 
very reasonable indeed.

\subsection{Photon-Photon Scattering}
Now, a typical T-matrix of the box diagrams in the photon-photon 
scattering can be written as
$$ T_{\gamma-\gamma} \simeq e^4 \int {d^4p \over (2\pi)^4}
{\rm Tr}\left[ (\gamma \epsilon_1){1\over \Slash{p}-\Slash{k}_1-\Slash{k}_2-m} (\gamma \epsilon_3)
{1\over \Slash{p}-\Slash{k}_3-m } (\gamma \epsilon_4) {1\over \Slash{p}-m}
 (\gamma \epsilon_2) {1\over \Slash{p}-\Slash{k}_2-m} \right] 
 \eqno{(2.3)}  $$
where the energy of photon can be written as 
$\omega=|\bm{k}_1|=|\bm{k}_2|=|\bm{k}_3|=|\bm{k}_4| $ at the center of mass system of 
two photons. The leading behavior of the finite terms in this T-matrix can be easily 
evaluated under the condition of $ m>> \omega $ as 
$$ T_{\gamma-\gamma} \sim {\alpha^2}\left[1  +c_1 \left({\omega\over m}\right)^2 
+c_2 \left({\omega\over m}\right)^4+..\right]  \eqno{(2.4)}  $$
where $c_1$ and $c_2$ denote some numerical constants. 
It should be important to note that the apparent divergences can be completely cancelled 
out due to the kinematical cancellation by adding up three independent Feynman diagrams 
together, and the disappearance of the divergences is not due to 
the regularization \cite{kanda}. As can be seen from eq.(2.4), the photon-photon 
cross section of Heisenberg-Euler in eq.(1.2) can be reproduced when one picks up 
the third term  ($c_2$ term) in eq.(2.4), and this is very strange from the point of view 
of the Feynman diagram evaluation. 

In terms of the reaction process, the total energy scale of the photon-photon 
cross section should be given by $2\omega$ which, in fact, appears in the denominator 
of eq.(1.1). Therefore, eq.(1.1) can be well understood since the T-matrix is given as 
eq.(2.4) at the low energy limit. 
On the other hand, the photon-photon cross section of eq.(1.2) can be obtained 
only when one picks up the third terms in eq.(2.4). 
The physical reason as to why eq.(1.2) cannot be justified is well explained in detail 
in \cite{kanda} in terms of the modern field theory terminology. Basically, one can see 
that the expression of eq.(1.2) can be obtained only when one picks up the smallest 
piece in eq.(2.4), and this is obtained only when one requires some unphysical 
conditions for the box diagrams in terms of the gauge conditions \cite{fujita}, which 
will be clarified in Appendix A.

\section{Possible Experiments}
If the photon-photon cross section of eq.(1.2) were correct, then there was no chance to 
observe it in terms of the laser-laser scattering experiment since the energy of the laser 
is mostly lower than a few tens of eV. However, the photon-photon scattering 
experiment at high energy must have a serious difficulty since the experiment 
should be done as the head-on collision in the center of mass system of two photons, and 
the control of the high energy photon flux must be non-trivially difficult. 

On the other hand, the situation is completely different if one should observe 
the photon-photon cross section of eq.(1.1). In this case, 
we should consider the laser-laser scattering where the energy of the laser 
is around $\omega =1$ eV or lower. If the measurement is carried out at 
$\theta =90$ degree, then the cross section of eq.(1.1) at $\omega \simeq 1$ eV becomes 
$$ {d\sigma\over d\Omega} \simeq {3\alpha^4\over 
(12\pi)^2 \omega^2} \simeq 2.3 \times 10^{-21} \ \ {\rm cm}^2 
\simeq 2.3 \times 10^{6} \ \ {\rm mb/st}  \eqno{(3.1)}    $$
which should be well detectable. 
On the other hand, the cross section of eq.(1.2) at $\omega \simeq 1$ eV becomes 
$$ {d\sigma\over d\Omega}\simeq {3\times417\alpha^4\over (180\pi)^2 m^2} 
\left({\omega\over m} \right)^6 
\simeq 9.3 \times 10^{-67} \ \ {\rm cm}^2 
\simeq 9.3 \times 10^{-40} \ \ {\rm mb/st}  \eqno{(3.2)}    $$
which is extremely small, and it is impossible to detect in any of the experiments. 
We note that the above cross section of eq.(1.2) becomes larger for larger photon energy. 
In fact, the cross section at $\omega \simeq 1$ MeV  becomes 
$$ {d\sigma\over d\Omega}\simeq {417\alpha^4\over (180\pi)^2 m^2} 
\left({\omega\over m} \right)^6 \simeq  9.3 \times 10^{-31} \ \ {\rm cm}^2 
\simeq 9.3 \times 10^{-4} \ \ {\rm mb/st} . \eqno{(3.3)}    $$
However, in this energy region, the low energy approximation is not well 
satisfied even though, we believe, the order of magnitude estimation must be correct. 
Unfortunately, there is no such high energy laser available at present.

\subsection{Comparison with $e^+ +e^- \rightarrow e^+ +e^- $ Scattering}
In terms of the reaction process, the photon-photon scattering must be similar 
to the $e^+e^- $ elastic scattering. The cross section of the  $e^+e^- $ elastic 
scattering process at high energy limit is given as
$$ {d\sigma\over d\Omega}\simeq {\alpha^2\over 8 E^2} 
\left[ {1+\cos^4(\theta /2)\over \sin^4(\theta /2) }  +
{1+\cos^2\theta \over 2} -{2\cos^4(\theta /2)\over \sin^2 \theta /2) }     \right]
. \eqno{(3.4)}    $$
The typical number of this cross section can be seen from experiment 
at $E=17$ GeV and $\theta =90$ degree, and it becomes 
$ {d\sigma\over d\Omega} \simeq 1.6 \times 10^{-7} \ \ {\rm mb/st} $. 
In order to find a naive estimation of the cross section at around $E\sim 2$ MeV, 
we extrapolate the energy dependence of eq.(3.4), and thus we find 
$ {d\sigma\over d\Omega} \sim 12 \ \ {\rm mb/st}  $. 
On the other hand, the photon-photon cross section of the present estimation 
at $\omega \sim 2$ MeV becomes 
$ {d\sigma\over d\Omega} \sim 6 \times 10^{-7} \ \ {\rm mb/st}  $ 
which is smaller than the  $e^+e^- $ elastic scattering cross section by 7 order of 
magnitudes. This naive estimation of the photon-photon cross section indicates 
that the cross section becomes quite large at low energy. However, in comparison 
with the head-on collisions between the  $e^+e^- $ elastic scattering cross section and 
the photon-photon cross section at 1 MeV incident energy, the photon-photon 
cross section is smaller than the $e^+e^- $ elastic scattering cross section by 
several orders of magnitude. 

\subsection{Comparison with $\gamma + \gamma \rightarrow e^++e^- $ Scattering}
If the energy of photon is larger than a few MeV, then we have to consider 
the scattering process in which the photon-photon scattering can produce 
the electron positron pair, that is, $\gamma + \gamma \rightarrow e^++e^- $. 
This cross section is the same order as the 
$e^+e^- $ elastic scattering cross section, and therefore, at higher energy 
than a few MeV, the photon-photon scattering process must be dominated by the 
$\gamma + \gamma \rightarrow e^+ +e^- $ cross section. 

In this respect, the $\gamma + \gamma \rightarrow e^++e^- $ should be used as 
the monitor of the reaction process before carrying out the photon-photon 
elastic scattering. This is clear since the main difficulty of the photon-photon 
scattering should be concerned with the initial conditions of photon-photon reaction, 
and therefore one should examine the validity of the reaction process first by carrying 
out the $\gamma + \gamma \rightarrow e^++e^- $ experiment. It should be noted that 
the photon-photon elastic cross section must be smaller than the $\gamma + \gamma 
\rightarrow e^++e^- $ reaction cross section by several orders of magnitudes. 
However, we believe it should be observed as long as we can judge from the magnitude of 
the cross section. 

\section{Discussions}
The real photon-photon cross section is much larger than the old expression 
which was obtained in the classical picture of the field theory. 
However, one can see that the reaction process can be observed only when 
the scattering should occur. The cross section we discuss is related to 
the probability of the scattering process when two photons collide. 
The basic difficulty of this scattering problem is indeed related to the fact 
that this scattering process is only possible for the head-on collision. 
Namely, the initial condition of the scattering process must be most difficult 
when setting up the photon-photon scattering experiment. 
In the case of Compton scattering, photon can scatter with electrons which are almost 
at rest while photon can interact with another photon only at the head-on collision 
since there is no photon at rest. In addition, both photons collide with each other 
at the speed of light. In this case, the focusing procedure must be made only 
in terms of the mechanical tools, in contrast to the $e^+e^-$ scattering 
process in which electrons can be controlled by the magnetic fields. 

What should be any realistic effects of the photon-photon scattering 
in nature ? It is most likely true that the possibility 
of photon-photon head-on collision which may happen in nature 
must be extremely small, and the only possible example may be found in the center 
of star where very many photons are created during the nuclear fusion processes, and 
these photons may collide with each other.


\newpage
\appendix

\section{Comments on Gauge Invariance}
There is a serious misunderstanding among some of the educated physicists concerning 
the gauge invariance of the calculated amplitudes which involve the external 
photon lines. Their argument is as follows. The polarization vector $\epsilon^\mu$ 
is gauge dependent and therefore the calculated results must be kept 
invariant under the transformation of $\epsilon^\mu \rightarrow \epsilon^\mu  + 
c k^\mu $. However, this condition is unphysical since we already fixed a gauge 
(for example, Lorentz gauge fixing of $ k_\mu \epsilon^\mu =0$) before the field 
quantization. The gauge invariance of the S-matrix evaluation is guaranteed 
as far as the fermion current is conserved, which is always satisfied in the perturbation  
calculation. Therefore, we briefly explain the gauge conditions by presenting a few 
examples. 

\subsection{Vacuum Polarization Tensor}
The best example can be found in the vacuum polarization tensor 
$\Pi^{\mu \nu}$. People believe that the following gauge condition should be satisfied
$$ k_\mu \Pi^{\mu \nu} =0  \eqno{(A.1)}  $$
which is required from the above argument of the gauge condition as well as some 
incorrect mathematical identity equation where the mistake is simply due to the wrong 
replacement of the integration variables in the infinite integrals. However, as one can 
easily examine it, this is a wrong equation \cite{fk}. In fact we can write the result 
of the standard calculation of the vacuum polarization tensor as
$$ \Pi^{\mu \nu}(k)=ie^2\int {d^4p\over(2\pi)^4}
{\rm Tr} \left[ \gamma^\mu {1\over p \llap/-m +i\varepsilon } \gamma^\nu  
{1\over p \llap/-k \llap/-m+i\varepsilon  }\right] = \ 
  {\alpha \over 2\pi} \left(\Lambda^2+m^2-{k^2\over 6} \right)g^{\mu \nu} $$
$$ +{\alpha \over 3\pi}(k^\mu k^\nu- k^2g^{\mu \nu})  
\left[ \ln \left({\Lambda^2\over m^2e}\right) -6\int_0^1 dz z(1-z) 
\ln \left(1-{k^2\over m^2}z(1-z) \right) \right]  \eqno{(A.2)}  $$
where $\Lambda$ denotes the cutoff momentum. There is no way that 
the first term of the right hand side can satisfy the gauge condition of eq.(A.1) 
\cite{fk,fujita}. 

Since then, however, people impose the gauge conditions by hand on the amplitudes 
which have some external photon lines. The gauge condition of $\epsilon^\mu \rightarrow 
\epsilon^\mu  + c k^\mu $ is not based on the solid physical principle, and therefore 
one can only say that, in some cases, the gauge condition can be satisfied.  Among the 
reactions that favour the gauge condition, there are two cases, the Compton scattering 
and  $\pi^0 \rightarrow 2 \gamma$. But other reaction processes cannot satisfy the gauge 
condition. 

\subsection{Compton Scattering}

As is well known, the Compton scattering satisfies the gauge condition, and this is 
mainly because there is no loop in this reaction and the external fermion line 
can satisfy the free Dirac equation. 
The Feynman amplitude of the Compton scattering can be written as
$$ {\cal M}^{\mu \nu}=-{ie^2}\left[ \bar{u}(p') \gamma^\nu 
{1\over p \llap/+k \llap/-m +i\varepsilon } \gamma^\mu u(p) 
+\bar{u}(p') \gamma^\mu {1\over p \llap/-{k \llap/}'-m +i\varepsilon } \gamma^\nu u(p) 
\right] \eqno{(A.3)}  $$
Therefore, we can check 
$$ k_\mu {\cal M}^{\mu \nu} =-{ie^2}\left[ \bar{u}(p') \gamma^\nu 
{1\over p \llap/+k \llap/-m +i\varepsilon } k \llap/ u(p) 
+\bar{u}(p') k \llap/ {1\over p \llap/-{k \llap/}'-m +i\varepsilon } \gamma^\nu u(p) 
\right]   .  \eqno{(A.4)}     $$
Now, using some identities 
$$ k \llap/ = p \llap/+k \llap/-m +(p \llap/-m), \ \ \ \ k \llap/ = 
-(p \llap/-{k \llap/}'-m) + ({p \llap/}'-m)  $$
and the Dirac equations of 
$$ (p \llap/-m) u(p)=0, \ \ \ \ \bar{u}(p') ({p \llap/}'-m)  =0 $$
we can easily prove 
$$ k_\mu {\cal M}^{\mu \nu}=-{ie^2}\left[ \bar{u}(p')\gamma^\nu u(p)
- \bar{u}(p')\gamma^\nu u(p)\right]  =0  .  \eqno{(A.5)}   $$
Therefore, the Compton scattering can satisfy the gauge condition, but this is, 
of course, clear since the diagram contains no loop. Thus, the gauge condition,  
$\epsilon^\mu \rightarrow \epsilon^\mu  + c k^\mu $ just corresponds to the conservation 
of the fermion current which is guaranteed by the free Dirac equation, and this can be 
easily seen since the initial and final fermion in the Compton scattering can satisfy 
the free Dirac equation. On the other hand, if the Feynman diagrams involve the fermion 
loop, then there is no reason that the gauge condition can directly correspond to 
the current conservation of fermions. 

\subsection{Decay of $\pi^0 \rightarrow 2  \gamma $}
In addition to the Compton scattering, the amplitude of $\pi^0 \rightarrow 2 \gamma $ 
decay can satisfy the above type of the gauge condition since it can be written as
$$ {\cal M}^{\mu \nu} ={\pi ge^2\over M} \varepsilon^{\mu \nu \rho \sigma} 
{k}_\rho {k'}_\sigma  .  \eqno{(A.6)} $$
In this case, it is easy to prove that 
$$ k_\mu {\cal M}^{\mu \nu} = {\pi ge^2\over M} \varepsilon^{\mu \nu \rho \sigma} 
k_\mu {k}_\rho {k'}_\sigma  =0  \eqno{(A.7)} $$
which is due to the anti-symmetric character of the $\varepsilon^{\mu \nu \rho \sigma}$ 
tensor. This property is basically due to the $\gamma_5$ interaction which generates 
the anti-symmetric nature of the invariant amplitude. In this respect, it is very 
special that the  $\pi^0 \rightarrow 2  \gamma $ decay process satisfies the gauge 
condition even though it has a fermion loop. However, it is not due to the nature of 
the electromagnetic interactions. In fact, this point can be clearly seen if we examine 
the following reaction process of the scalar meson decay into two photons which cannot 
satisfy the gauge condition. 

\subsection{Decay of Scalar Boson $\Phi$ into $ 2  \gamma $}
Now, the T-matrix of $\Phi \rightarrow 2  \gamma $ which is based on the triangle diagrams 
can be evaluated to be
$$ T_{\Phi\rightarrow 2 \gamma} \simeq e^2 g_0 \int {d^4p\over (2\pi)^4} {\rm Tr} 
\left[(\gamma \epsilon) {1\over p \llap/-M +i\varepsilon } 
(\gamma {\epsilon}') {1\over p \llap/+{k \llap/}-M +i\varepsilon } 
{1\over p \llap/-{k \llap/}'-M+i\varepsilon  }  \right] 
\simeq { e^2g_0} M (\epsilon {\epsilon}')   \eqno{(A.8)} $$
where $M$ denotes the nucleon mass, and $g_0$ is the coupling constant of $\Phi N$ 
interaction as described by ${\cal L}_{II}=g_0\bar{\psi} \psi \Phi $. 
Defining the amplitude ${\cal M}^{\mu \nu}$ as $T_{\Phi \rightarrow 2 \gamma}=
{\cal M}^{\mu \nu} \epsilon_{\mu} {\epsilon'}_{\nu} $, we can now prove  
$$ k_\mu {\cal M}^{\mu \nu} = { e^2g_0} M k_\mu g^{\mu \nu} \not= 0 . \eqno{(A.9)} $$
Therefore, the gauge condition is not satisfied in the case of 
of $\Phi \rightarrow 2  \gamma $ decay process. This is, of course, clear 
since the scalar interaction has a symmetric nature and therefore it is just opposite 
to the $\gamma_5$ interaction. 
It should be noted that there is no scalar meson in nature which decays into two photons. 
However, the similar type of the Feynman diagram becomes important when 
we consider the photon-gravity interaction. In fact, photon can interact 
with the gravitational field via loop diagrams which are essentially the same as
the T-matrix given in eq.(A.8) \cite{fujita}. In this respect, the T-matrix 
given in eq.(A.8) can be considered to be a real physical process. 

\subsection{Summary of Gauge Conditions}
To summarize, we see that the amplitudes of $\pi^0 \rightarrow 2  \gamma $ and 
the Compton scattering happen to satisfy the gauge condition, while other examples of 
the photon self-energy, scalar meson decay into two photons and photon-photon 
scattering diagrams do not satisfy the gauge condition. The  $\pi^0 \rightarrow 2 
 \gamma $ case satisfies the gauge condition due to the anti-symmetric nature of 
the pion-nucleon interactions while the Compton scattering diagrams can satisfy 
the gauge condition because they contain no fermion loops. 

In general, the problem of the gauge conditions can be better understood if we give 
one example in terms of the Lorentz invariance. Now, the S-matrix in QED is formulated 
in a covariant fashion, and therefore the cross section defined from 
the T-matrix is, for sure, Lorentz invariant. When calculating the cross 
section, one should choose the system such as the laboratory system or the center of 
mass system, depending on the experimental or theoretical situations. Then, one can 
evaluate the cross section and can reliably obtain the calculated result since the Lorentz 
invariant quantity can be calculated at any system one wishes. At this point, a physics 
student may ask a question as to what should be the Lorentz invariance of the cross 
section. Of course, one knows that this is a meaningless question since the calculation 
is done by fixing the system. However, as one sees by now, the requirement of 
the gauge condition of $\epsilon^\mu \rightarrow \epsilon^\mu  + c k^\mu $  is 
just the same level as the above question.


\begin{thebibliography}{99}

\bibitem{kanda}
N. Kanda, "Light-Light Scattering", to be published

\bibitem{landau}
V.B. Berestetskii, E.M. Lifshitz and L.P. Pitaevskii, "Relativistic Quantum Theory" 
(Pergamon Press, 1974) 

\bibitem{euler}
W. Heisenberg and H. Euler, Z. Phys. {\bf 98}, 714 (1936)

\bibitem{karp}
R. Karplus and  M. Neuman, Phys. Rev. {\bf 83},776 (1951) 


\bibitem{fk}
T. Fujita and N. Kanda, "Tomonaga's Conjecture on Photon Self-Energy", 
physics.gen-ph/1102.2974


\bibitem{exp1}
F. Moulin, D. Bernard, and F. Amiranoff, Z. Phys. {\bf C 72}, 607 (1996)

\bibitem{exp2}
D. Bernard, F. Moulin, F. Amirano, A. Braun4, J.P. Chambaret, G. Darpentigny, 
G. Grillon, S. Ranc, and F. Perrone, Eur. Phys. J. {\bf D 10}, 141 (2000)




\bibitem{nishijima}
K. Nishijima, " Fields and Particles " , (W.A. Benjamin,INC) 

\bibitem{fujita}
T. Fujita, "Symmetry and Its Breaking in Quantum Field Theory", 
(2nd edition, Nova Science Publishers, 2011) 


\end{thebibliography}
\end{document}